\documentclass{elsart}

\usepackage{amssymb,graphics}

\begin{document}

\begin{frontmatter}

\title{Incipient 
quantum melting of the one--dimensional Wigner 
   lattice}

\author[A]{S. Fratini},
\author[B]{B. Valenzuela} and 
\author[C]{D. Baeriswyl}
\address[A]{Laboratoire d'Etudes des 
Propri\'et\'es Electroniques
  des Solides, CNRS, 25 avenue des 
Martyrs, BP\ 166,
F-38042 Grenoble Cedex 9, 
France}
\address[B]{Instituto de Ciencia de Materiales, 
CSIC,
Cantoblanco, E-28049 Madrid, Spain}
\address[C]{D\'epartement 
de Physique, Universit\'e de
  Fribourg, P\'erolles,
CH-1700 
Fribourg, Switzerland}

\begin{abstract}
A one--dimensional 
tight--binding model of electrons with long--range 
Coulomb
interactions is studied in the limit where double site 
occupancy is forbidden
and the Coulomb coupling strength $V$ is large 
with respect to the hopping amplitude
$t$. The quantum problem of a 
kink--antikink pair generated in the Wigner lattice
(the classical 
ground state for $t=0$) is solved for fillings $n=1/s$, where $s$
is 
an integer larger than 1. The pair energy becomes negative for a 
relatively 
high value of $V$, $V_c/t\approx s^3$. This signals the 
initial stage of the quantum 
melting of the Wigner 
lattice.
\end{abstract}

\end{frontmatter}

\section{Introduction}
Three decades ago, Michael Rice and collaborators have
    introduced nonlinear phase excitations of a charge--density--wave
    condensate as a new type of charged states and referred
    to them as ${\varphi}$ particles \cite{phiparticles}. Their arguments were
    based on the Peierls instability of a one--dimensional coupled
    electron--phonon system, in the limit where the
    charge--density wave amplitude is much smaller than the
    average electronic charge density. Shortly after, Hubbard
    discussed qualitatively the role of dimer pairs (domain
    boundaries) in a nearly quarter--filled band with dominant
    long--range Coulomb interactions \cite{hubbard}. As we will argue, the
    two types of charged states are closely related.

We consider a one--dimensional system of spinless fermions
described by the Hamiltonian
\begin{equation}
H=-t\sum_i(c_i^{\dagger}c_{i+1}+c_{i+1}^{\dagger}c_{i})
+\sum_i\sum_{l\geq 1}V_ln_in_{i+l}\ ,
\label{hamiltonian}
\end{equation}
where $c_i^{\dagger}$ and $c_i$ are fermionic creation and
annihilation operators, respectively, $n_i=c_i^{\dagger}c_i$ is
the occupation number of site $i$ and $V_l=V/l$ represents
the long--range Coulomb potential.
Such a model may describe molecular chain compounds where the
interaction between two (valence) electrons on the same
molecule is so large that double occupancy can safely be discarded.
The spin quantum number is then redundant because the exchange
of two electrons is dynamically forbidden.

We have in mind systems where the average site occupation is a
rational number between 0 and 1, $n=r/s$. Depending on the band
filling the fermions then represent either electrons or holes.
The classical ground state ($t=0$) is charge--ordered and forms a
``generalized Wigner lattice'' \cite{hubbard}. In the special
case $n=1/s$ the unit cell contains $s$ sites, one of which is
occupied. A generalized Wigner lattice is clearly insulating and
therefore the long--range Coulomb interactions remain unscreened.
Insulating charge--ordered phases have been reported for organic
chain compounds \cite{brazovskii} and in several cases the nature
of charge ordering is more likely that of a Wigner crystal
than that of a (small amplitude) charge--density wave.

We have recently used a variational wave function for describing
the effects of a small but finite hopping term ($t \ll V$)
\cite{belen}. Here
we consider the same regime, but follow a different route by
examining the stability of the classical ground state (the
generalized Wigner lattice) with respect to the spontaneous generation
of charge defects, as was done in ref. \cite{quemerais} in the case of
electron-phonon interactions.
Such defects occur in pairs (kinks and antikinks)
and can be viewed as fractionally charged particles, in close
analogy to the ${\varphi}$ particles of Michael Rice.

\section{Kink--antikink pair for $n=1/2$}
We start our discussion for a density $n=1/2$. In order to guarantee overall
charge neutrality we introduce a rigid compensating background ($n=1/2$ charge
of opposite sign at each site). The classical ground state in this
case corresponds to alternating filled and empty sites (Fig.\ 1a). Note
that this configuration, already at the classical level, is doubly
degenerate, since exchanging the empty and occupied sites does not change
the overall energy. The most simple defect in this perfect structure, called
{\it kink} in the following, is a domain boundary separating the two possible
ground state configurations. It can be visualized as  a pair of
occupied nearest neighbor sites (a dimer) and is necessarily accompanied
by an {\it antikink}, a pair of empty nearest--neighbor sites.
\begin{figure}
\centerline{\includegraphics{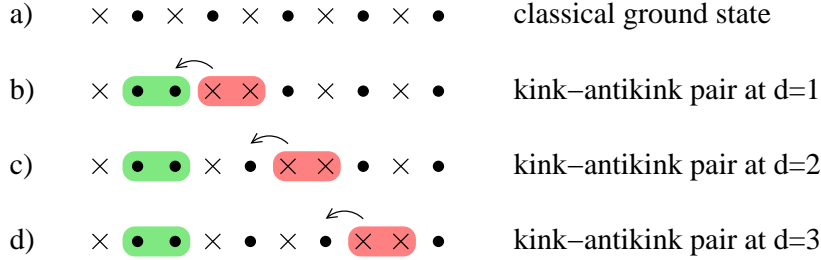}\\ \medskip }
\caption{Classical ground state and low-energy defect pairs at $n=1/2$. 
Dots/crosses stand for filled/empty sites respectively. The kinks and 
anti-kinks correspond to the shaded areas. The arrows indicate 
the hopping processes leading to a given configuration.}
\end{figure}

The classical energy of a kink--antikink pair is readily evaluated.
Introducing a pair of size $2d$, {\it i.e.}\ a kink and an antikink
separated by a distance $2d$ (see Fig.\ 1), costs an energy
\begin{equation}
   \label{eq:kk-energy-cl}
   \Delta (d)=\sum_{p=1}^d \sum_{m=p}^\infty [V_{2m+1}+V_{2m-1}-2V_{2m}] .
\end{equation}
This sum can be transformed into a power series in $1/(2d)$ with
leading terms
\begin{equation}
   \label{eq:kk-approx}
\Delta (d)=\frac{V}{2}-\frac{V}{4}\frac{1}{2d}+\frac{V}{8}\frac{1}{(2d)^3}\ .
\end{equation}
Here the first term represents the creation energy of two well separated
defects. The second term is the Coulomb attraction between kink and antikink
with a coupling reduced by a factor of 4 as compared to the original
Hamiltonian (\ref{hamiltonian}). Therefore the defects carry effective
charges $\pm 1/2$ (measured in units of the electronic charge). This result
agrees with what one obtains by adding or subtracting an electron to the
classical ground state configuration \cite{hubbard}. In fact, adding an
electron at an empty site results in a high--energy configuration with
a three--electron cluster. The energy is lowered by moving away one of
these electrons, thus creating a pair of defects (two kinks in our
language). Since the overall added charge is -1, each kink carries a
charge -1/2. Similarly, removing an electron produces
two antikinks, each of which carries a charge +1/2. The third term in
Eq.\ \ref{eq:kk-approx} can be interpreted as the interaction of two
dipole moments pointing in opposite directions parallel to the chain axis.
The size of the dipoles is equal to the fractional charge times half a lattice
constant, in our units $1/4$. We conclude that kinks and antikinks behave
like fractionally charged particles with electric dipole moments. The
dominant interaction is the Coulomb attraction, even at the shortest possible
distance $2d=2$.

We now turn to the quantum problem of a defect pair, {\it i.e.}\ we
diagonalize the Hamiltonian within the subspace of states $|m,d\rangle$,
corresponding to a kink at site $m-d$ and an antikink at site $m+d$ ($m$
thus indicates the center-of-mass of the pair). In general the
effect of a hopping event on the state $|m,d\rangle$ is either to
move an existing defect by two lattice sites, which
modifies both the size ($2d\to 2d\pm 2$) and the center ($m\to m=\pm
1$) of the pair (see Fig.\ 1), or to create (annihilate) an additional pair.
Restricting ourselves to the subspace of
single--pair states, creation and annihilation processes are forbidden and
the quantum problem reduces to the eigenvalue equation

\begin{eqnarray}
\left[H - \Delta (d)\right] |m,d\rangle
  + t\   [  & & |m-1,d+1\rangle +|m+1,d-1\rangle \nonumber\\
   + &&  |m+1,d+1\rangle +|m-1,d-1\rangle ] =0\ .
   \label{eq:pair-quantum}
\end{eqnarray}
The center--of--mass and relative motions can be separated by
introducing the Bloch superposition
\begin{equation}
   \label{eq:psi}
   |\psi\rangle =\sum_{m,d\ge 1} e^{i K m} \psi (d) |m,d\rangle\ .
\end{equation}
Introducing this Ansatz
into Eq.\ (\ref{eq:pair-quantum}) leads to the following
eigenvalue equation for the wave function $\psi(d)$,
\begin{equation}
   \label{eq:eigenvalue}
   E \psi(d)=\Delta  (d)  \psi(d) - 2 t \cos K
   [\psi(d+1)+\psi(d-1) ]\ , \;\;  d\ge 1\ .
\end{equation}
Seeking for the lowest--energy wave function we restrict ourselves to the pair
at rest ($K=0$). The classical ground state (no defect pairs)
is excluded by imposing the boundary condition $\psi(0)=0$.
This eigenvalue problem can be treated to arbitrary accuracy numerically, but
more insight is gained by solving it in the continuum limit,
$\psi(d)\rightarrow\psi(x)$, $\psi(d\pm1)\rightarrow \psi(x\pm 2)$,
where $x$ is the position in units of the distance between neighboring sites.
Neglecting the
dipolar interaction in Eq.\ (\ref{eq:kk-approx}) we obtain the eigenvalue
equation
\begin{equation}
   \label{eq:eigenv-continuum}
   \epsilon \psi(x)=-8t \frac{d^2 \psi}{dx^2}- \frac{V}{4x}\psi(x)\ ,
\end{equation}
where $\epsilon=E-V/2+ 4t$.
Eq.\ (\ref{eq:eigenv-continuum}) is equal to the radial eigenvalue equation of
the hydrogen problem in the s-wave channel, and the boundary conditions
are also the same, $\psi(0)=0$. Therefore we can immediately write down the
lowest--energy wave function of a pair,
\begin{equation}
   \label{eq:wf-continuum}
   \psi_0(x)=2 a_0^{-3/2} x e^{-x/a_0},\ x\ge 0\ ,
\end{equation}
with an effective Bohr radius
\begin{equation}
a_0 = \frac{64t}{V}
\label{bohr-radius}
\end{equation}
and an energy
\begin{equation}
E_0=-4t+\frac{V}{2}-\frac{V^2}{512 t}\ .
\label{eq:epair}
\end{equation}
When the interaction strength is lower than the critical
value $V_c/t=8.27$, $E_0<0$ and the
formation of a kink-antikink pair becomes favorable by virtue of the gain in
energy achieved through delocalization as a Bloch state.
\footnote{The continuum description is appropriate provided that
the extent $\langle x\rangle=96 t/V$ is larger than the average
distance between particles, a
condition that is well satisfied at the critical
coupling $V_c$. Indeed, solving
the original discrete problem of Eq.\ (\ref{eq:eigenvalue}) numerically yields
$V_c/t=8.25$, which is very close to the continuum estimate.}

At the critical interaction strength the Wigner lattice
will be destabilized due to a proliferation of kink--antikink pairs, but
we do not expect a true transition to a one--dimensional metallic state
(a Luttinger liquid), because for arbitrarily small $V$ the system is
unstable with respect to a charge--density wave with the same period as
for the Wigner lattice \cite{schulz}. Therefore the critical strength $V_c$
indicates a crossover between a Wigner lattice and small--amplitude
charge--density wave. Note that in the crossover region the binding energy
$V^2/512 t$ makes a
negligible correction to the quantum mechanical energy of a pair: the
main contributions to Eq.\ (\ref{eq:epair}) are
the delocalization energy
$-4t$, and the formation energy of the isolated defects $V/2$.

\section{Kink--antikink pairs for $n=1/s$}

We extend now the considerations
of the previous section to other filling factors $n\neq 1/2$. For a general
value of $n$ already the classical limit ($t=0$) is quite involved, leading to
complicated charge patterns \cite{hubbard}. Therefore it is worthwhile to
make contact to the opposite limit, $V<t$, where a simple picture is available.
Mean--field theory applied to the Hamiltonian (1) yields a ground state with
a modulated charge density
\begin{equation}
\langle\delta n_l\rangle = a\ cos(Ql+\varphi_l)\ ,
\label{cdw}
\end{equation}
where $Q=2k_F=2\pi n$ in the case of spinless fermions, and $a\ll n$ for
$V<t$. The phase $\varphi_l$ is an arbitrary constant for an incommensurate
situation ($n$ irrational), but locked to one of $s$ possible values in the
commensurate case, where $n=r/s$ with integer numbers $r$ and $s$
\cite{gruner}.
Adding an electron to an
incommensurate charge--density wave results in a slight shift of the wave
vector $Q$, $Q\rightarrow [1+(1/L)]Q$, where $L$ is the chain length.
In view of Eq.\ (\ref{cdw}), this effect can also
be attributed to a phase shift $\varphi_l=2\pi l/L$. For a
commensurate case the phase achieves the change of $2\pi$ from $l=0$ to $l=L$
through $s$ steps, {\it i.e.}\ there are $s$ $\varphi$--particles, each of
which
carries a fractional charge $-1/s$.

\begin{figure}
\centerline{\includegraphics{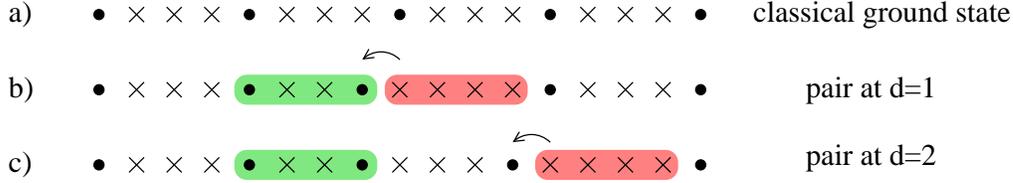} \\ \medskip}
\caption{Classical ground state and low energy defect pairs at $n=1/4$. 
The symbols are the same as in figure 1.}
\end{figure}

Let us now return to the large $V$ limit and restrict ourselves
to simple ratios $n=1/s, s\ge 3$. In this case the classical ground state
is $s$--fold degenerate and there are several types of domain walls
separating the different configurations.
It turns out that the low--energy domain walls are those which connect
nearby ground state configurations, {\it i.e.}\ Wigner lattices where
the locations of electrons differ by one lattice constant (see Fig.\ 2).
These ``kink'' and ``antikink'' defects again occur in pairs.
Repeating the arguments of the steps of the previous section, we evaluate
the classical energy of a kink--antikink pair of size $sd$,
\begin{equation}
   \label{eq:kk-energy-general}
   \Delta(d)=\sum_{p=1}^d \sum_{m=p}^\infty
   \left(V_{sm+1}+V_{sm-1}-2V_{sm}\right).
\end{equation}

\noindent
The summation can again be performed for large sizes, and we find up to first
order in $1/(sd)$
\begin{equation}
\Delta(d)=\frac{V}{s}\left[1-\frac{\pi}{s}\cot
\left(\frac{\pi}{s}\right)\right]
-\frac{V}{s^2}\frac{1}{sd}\ .
\label{s-energy}
\end{equation}
\noindent
The first term, the pair creation energy $\Delta_\infty$ at infinite
separation,
agrees with the corresponding quantity in Eq.\ (\ref{eq:kk-approx}) for $s=2$.
The second term corresponds to the Coulomb attraction for two particles with
fractional charges $\pm1/s$ at a distance $sd$, in agreement with the counting
argument presented above for the $\varphi$--particles.

We can again use the continuum limit for calculating the lowest--energy
quantum pair state for general $s$. The wave function is again given
by Eq.\ (\ref{eq:wf-continuum}), but with a Bohr radius
\begin{equation}
a_0=\frac{4s^4t}{V}\ .
\end{equation}
Thus the pair size increases strongly with decreasing density $n=1/s$.
Correspondingly, the binding energy decreases, as seen in the pair
energy
\begin{equation}
E_0=-4t+\Delta_\infty-\frac{V^2}{8s^6t}\ ,
\end{equation}
where the first two terms dominate for large $s$. In fact, for $s\gg 1$
we can safely use the asymptotic value
\begin{equation}
E_0 \approx -4t+\frac{\pi^2}{3s^2}V
\end{equation}
to estimate the instability point, where this energy becomes negative.
The result,
\begin{equation}
\frac{V_c}{t}\approx\frac{12}{\pi^2n^3}
\end{equation}
agrees well with our previous variational estimate \cite{belen}. This is
not surprising because the creation of a pair can be achieved
through a hopping event that moves a particle out of the ground state
configuration. The probability of such a hopping event was evaluated in
Ref.\ \cite{belen}, and was used to determine a criterion for the
instability of the generalized Wigner lattice.

In contrast to the case $n=1/2$, for lower density this instability
does not signal a crossover to a small--amplitude charge--density wave,
it rather indicates that part of the electronic charge is spilled
over to neighboring sites of the classical Wigner lattice.

\section{Discussion}

In this paper we have determined the lowest--energy quantum state of
a kink--antikink pair in a one--dimensional generalized Wigner lattice.
We have calculated the critical value $V_c$ of the interaction strength
below which the pair energy is negative, and charge defects will be
generated spontaneously. Kinks (or antikinks) are the 
strong--coupling
analogs of the $\varphi$--particles studied a long 
time ago by
Michael Rice and collaborators \cite{phiparticles}.

The true quantum ground state of electrons interacting through
long--range Coulomb forces contains kink--antikink pairs due to quantum
fluctuations, even above $V_c$. It would be interesting to proceed
from the single--pair solution to that of an arbitrary number of pairs.
This step is highly non--trivial, although at first sight it looks
similar to that from the Cooper problem to the BCS wave function. One
of the difficulties arises from the unknown statistics of kinks and antikinks,
another from their non--local character.

If the density of pairs in the ground state is large enough, they
loose their identity, and kinks and antikinks may move rather independently.
In this case one can imagine a d.c.\ charge transport due to moving defects.
Whether such a
mechanism is responsible for the observed Drude peak in the Bechgaard salts
\cite{degiorgi} is an interesting open question.

\section*{Acknowledgments}
One of us (S.F.) would like to thank P. Qu\'emerais for suggesting the
idea of defect pairs describing the melting of the classically ordered state.


\begin{thebibliography}{00}

\bibitem{phiparticles} M.\ J.\ Rice, A.\ R.\ Bishop, J.\ A.\ Krumhansl
and S.\ E.\ Trullinger, Phys.\ Rev.\ Lett.\ {\bf 36}, 432 (1976).
\bibitem{hubbard} J.\ Hubbard, Phys.\ Rev.\ B {\bf 17}, 494 (1978).
\bibitem{brazovskii} For a collection of recent results see the
{\it Proceedings of the International Workshop on Electronic Crystals},
edited by S.\ E.\ Brazovskii, N.\ Kirova and P.\ Monceau,
J.\ Phys.\ (France) IV {\bf 12}, Pr 9 (2002).
\bibitem{belen} B.\ Valenzuela, S.\ Fratini and D.\ Baeriswyl,
Phys.\ Rev.\ B {\bf 68}, 045112 (2003).
\bibitem{quemerais} S. Aubry and P. Qu\'emerais, in 
\textit{Singular Behaviour and nonlinear dynamics}, ed. St. Pnevmatikos, 
T. Bountis and Sp. Pnevmatikos, World Scientific, p. 342 (1989); P. Qu\'emerais, 
D. K. Campbell, J. L. Raimbault, S. Aubry, Int. Journ. of Mod. Phys. {\bf B 7}, 4289 (1993)
\bibitem{schulz} H.\ J.\ Schulz, Phys.\ Rev.\ Lett.\ {\bf 71}, 1864 (1993).
\bibitem{gruner} See for instance G.\ Gruner, {\it Density Waves in Solids},
Addison--Wesley, Reading (1994).
\bibitem{degiorgi} For a recent review  see L.\ Degiorgi,
in {\it Strong Interactions in Low Dimensions},
edited by D.\ Baeriswyl and L.\ Degiorgi, Kluwer Academic Publishers,
to appear.


\end{thebibliography}
\end{document}